\documentclass{emulateapj}
\usepackage{apjfonts}

\newcommand{\re}{r_{\rm E}}

\newcommand{\thetae}{\theta_{\rm E}}
\newcommand{\deltavec}{\mbox{\boldmath $\delta$}}
\newcommand{\zetavec}{\mbox{\boldmath $\zeta$}}

\shortauthors{HAN}
\shorttitle{FREE-FLOATING PLANETS}

\begin{document}

\title{Secure Identification of Free-Floating Planets}

\author{Cheongho Han}
\affil{Department of Physics, Institute for Basic Science
Research, Chungbuk National University, Chongju 361-763, Korea;\\
cheongho@astroph.chungbuk.ac.kr}


\begin{abstract}
Among the methods proposed to detect extrasolar planets, microlensing 
is the only technique that can detect free-floating planets. 
Free-floating planets are detected through the channel of short-duration 
isolated lensing events.  However, if a seemingly isolated planetary 
event is detected, it is difficult to firmly conclude that the event 
is caused by a free-floating planet because a wide-separation planet 
can also produce an isolated event.  There were several methods 
proposed to break the degeneracy between the isolated planetary 
events produced by the free-floating and wide-separation planets, 
but they are incomplete.  In this paper, we show that free-floating 
planets can be securely identified by conducting astrometric follow-up 
observations of isolated events to be detected in future photometric 
lensing surveys by using high-precision interferometers to be operated 
contemporarily with the photometric surveys.  The method is based on 
the fact that astrometric lensing effect covers much longer range of 
the lens-source separation than the photometric effect.  We demonstrate 
that several astrometric follow-up observations of isolated planetary 
events associated with source stars brighter than $V\sim 19$  by using 
the {\it Space Interferometry Mission} with an exposure time of 
$\lesssim 10~{\rm min}$ for each observation will make it possible 
to measure the centroid shift induced by primaries with projected 
separations up to $\sim 100~{\rm AU}$.  Therefore, the proposed 
method is far more complete than previously proposed methods that 
are flawed by the limited applicability only to planets with projected 
separations $\lesssim 20~{\rm AU}$ or planets accompanied by bright 
primaries. 
\end{abstract}


\section{Introduction}

Although originally proposed as a method to search for Galactic 
dark matter in the form of massive compact objects \citep{paczynski86},
microlensing has developed into important tools in various aspects 
of astrophysics (see the review \citealt{gould01}), including the 
detection and characterization of extrasolar planets \citep{mao91}.
Recently, three robust microlensing detections of exoplanets were 
reported by \citet{bond04}, \citet{udalski05}, and \citet{beaulieu06}.

The microlensing signal of a planet is a short-duration perturbation 
to the smooth standard light curve of the primary-induced lensing 
event occurring on a background source star.  The planetary lensing 
signal induced by a giant planet with a mass equivalent to that of 
the Jupiter lasts for a duration of $\sim 1$ day, and the duration 
decreases in proportion to the square root of the mass of a planet, 
reaching several hours for an Earth-mass planet.  To achieve the 
observational frequency required to detect the short-lived planetary 
signal, current microlensing planet search experiments are operated 
in a special observational setup, where survey observations issue 
alerts of ongoing events in the early stage of lensing magnification 
\citep{udalski94, alcock97, bond01} and follow-up observations 
intensively monitor the alerted events \citep{albrow98, yoo04}.
However, follow-up is generally done with an instrument having a 
small field of view, and thus events should be monitored in sequence.
As a result, only a handful number of events can be followed at any 
given time, and thus the number of planet detections is limited.

However, the situation will be different in next-generation lensing 
experiments that will survey wide fields continuously at high cadence 
by using very large-format imaging cameras.  Several such surveys in 
space and on the ground have already been proposed or are being 
seriously considered.  The {\it Galactic Exoplanet Survey Telescope} 
({\it GEST}), whose concept was succeeded by the {\it Microlensing 
Planet Finder} ({\it MPF}), is a space mission to be equipped with 
a 1--2 m aperture telescope \citep{bennett02, gould05}.  The 
`Earth-Hunter' project is a ground-based survey that plans to 
achieve $\sim 10$ minute sampling by using a network of three 2 m 
class wide-field ($\sim 2^\circ\times2^\circ$) telescopes scattered 
over the southern hemisphere (A.\ Gould, private communication).  
These next-generation surveys dispense with the alert/follow-up mode 
of searching for planets, and instead simultaneously obtain 
densely-sampled lightcurves of all microlensing events in their 
field-of-view, yielding improved sensitivity to planets.

In addition to the dramatically improved sensitivity to planets, 
the next-generation lensing surveys can detect two new populations 
of wide-separation and free-floating planets \citep{bennett02, han04, 
han05}.  In the current planet search strategy based on alert/follow-up 
mode, events can be followed only when the source star is located 
within the Einstein ring of the primary lens.  Such source positions 
are only sensitive to planets located within a certain range of 
distance from their host stars \citep{gould92}.  Planets in this 
so-called lensing zone have separations in the range of $0.6\lesssim 
s \lesssim 1.6$, where $s$ is the  projected star-planet separation 
normalized by the Einstein radius $r_{\rm E}$.  The Einstein radius 
is related to the mass of the lens (i.e., the host star of the planet) 
$m$ and distances to the lens $D_{\rm L}$ and source (background 
lensed star) $D_{\rm S}$ by
\begin{equation}
\re \simeq 4.9\ {\rm AU}\ \left( {m\over 0.5\ M_\odot}\right)^{1/2}
\left( {D_{\rm L}\over 6\ {\rm kpc}}\right)^{1/2}
\left( 1-{D_{\rm L}\over D_{\rm S}}\right)^{1/2}.
\label{eq1}
\end{equation}
For a typical Galactic bulge event caused by a low-mass stellar 
lens with $m=0.5\ M_\odot$, $D_{\rm L}=6\ {\rm kpc}$, and $D_{\rm S}
=8\ {\rm kpc}$, the Einstein radius is $r_{\rm E}\sim 2.5$ AU.  
Therefore, current microlensing planet search strategy is sensitive 
only to {\it bound} planets in the range of the projected separation 
of $1\ {\rm AU} \lesssim r_\perp\lesssim 5\ {\rm AU}$.  In the future 
survey, however, events will be monitored regardless of whether the 
primary of a planet magnifies the background source star or not, and 
thus wide-separation and free-floating planets can be detected through 
the channel of {\it isolated} events.

Among the two new populations of planets to be detected in the future 
lensing surveys, free-floating planets are of special interest.
Most theories of planet formation predict that a large number of 
planets are ejected from their planetary systems during or after 
the epoch of their formation.  However, it would be difficult to 
test these theories without a capacity to detect the free-floating 
planets arising from such ejections.  The importance of microlensing 
detections of free-floating planets lies in the fact that microlensing 
is the only proposed method that can detect these planets.  All other 
techniques rely on the effect of the planet on its parent star (either 
jostling the star's position or blocking its light), and thus they can 
detect only planets that are bound to their parent stars.

However, there is a problem in identifying free-floating planets.
The problem is that both the free-floating and wide-separation planets 
exhibit their existence through the same channel of isolated events. 
\citet{han05} proposed three methods for distinguishing isolated events 
caused by the two populations of planets.  These include detecting the 
primary through the influence of the planetary caustic, through the 
low-amplitude bump in the lightcurve from the primary, and through 
the detection of the light from the primary itself (see \S\ 3 for 
details).  However, a substantial fraction of events caused by 
wide-separation planets do not show symptoms of their primaries, and 
thus they cannot be distinguished from those produced by free-floating 
events.  As a result, if a seemingly isolated planetary event is 
detected, it is difficult to firmly conclude that the event is 
caused by a free-floating planet.

In addition to photometric observations, microlensing events can 
also be observed astrometrically by using future high-precision 
interferometers such as those to be mounted on space-based platforms, 
e.g.\ the {\it Space Interferometric Mission} ({\it SIM}) and the 
{\it Global Astrometric Interferometers for Astrophysics} ({\it GAIA}), 
and those to be mounted on very large ground-based telescopes, e.g.\ 
Keck and VLT.  If an event is astrometrically observed by using these 
instruments, it is possible to measure the lensing-induced positional 
displacement of the centroid of the source star image with respect to 
its unlensed position \citep{hog95, miyamoto95, walker95, paczynski98, 
boden98, han99}.  Recently, the proposal of astrometric microlensing 
observations was selected as one of the long-term projects of the 
{\it SIM}, which is scheduled to be launched in 2013.  Then, a 
significant time of the next-generation photometric lensing surveys 
will overlap with that of the {\it SIM} observations, and thus 
astrometric follow-up observations of events detected from the 
photometric surveys will be possible in the future.  In this paper, 
we demonstrate that astrometric follow-up observations of an isolated 
planetary lensing event by using high-precision interferometers can 
firmly distinguish whether the event is produced by a wide-separation 
planet or a free-floating planet.

The paper is organized as follows.  In \S\ 2, we briefly describe 
basics of photometric and astrometric behaviors of microlensing events.
In \S\ 3, we mention about the previously proposed methods to break 
the degeneracy between the isolated planetary lensing events caused by 
wide-separation and free-floating planets and discuss the limitations 
of the methods.  In \S\ 4, we describe the basic scheme of the 
proposed method and demonstrate the feasibility of the method.  
We also discuss the superiority of the proposed method over the 
previously proposed methods.  In \S\ 5, we summarize and conclude.

\section{Basics of Photometric and Astrometric Lensing}

A planetary lensing is described by the formalism of a binary 
lensing with a very low-mass companion.  For binary lensing, the 
mapping from the lens plane to the source plane is expressed as  
\begin{equation}
\zeta = z - \sum_{j=1}^2 {m_j/m \over \bar{z}-\bar{z}_{L,j}},
\label{eq2}
\end{equation}
where $\zeta=\xi + i\eta$, $z_{L,j}=x_{L,j}+iy_{L,j}$, and $z=x+iy$ 
are the complex notations of the source, lens, and image positions, 
respectively, $\bar{z}$ denotes the complex conjugate of $z$, $m_j$ 
are the masses of the individual lens components, and $m=m_1+m_2$ 
\citep{witt90}.  Here all angles are normalized to the angular size 
of the Einstein radius of the total mass of the system, i.e.\
\begin{equation}
\thetae = 
{\re\over D_{\rm L}} \simeq 710\ 
{\rm \mu as}\ \left( {m\over 0.5\ M_\odot}\right)^{1/2}
\left( {D_{\rm S}\over 8\ {\rm kpc}}\right)^{-1/2}
\left( {D_{\rm S}\over D_{\rm L}}-1\right)^{1/2}.
\label{eq3}
\end{equation}
The lensing process conserves the source surface brightness, and 
thus the magnifications $A_i$ of the individual images correspond 
to the ratios between the areas of the images and the source.  
For an infinitesimally small source element, the magnification is 
mathematically expressed as 
\begin{equation}
A_i = \left\vert \left( 1-{\partial\zeta\over\partial\bar{z}}
{\overline{\partial\zeta}\over\partial\bar{z}} \right)^{-1}
\right\vert.
\label{eq4}
\end{equation}
Then the total magnification is sum of the magnifications of the 
individual images, $A= \sum_i A_i$.  The position of the image 
centroid corresponds to the magnification weighted mean position 
of the individual images, and thus the displacement of the position 
of the source star image centroid with respect to its unlensed 
position (centroid shift) is expressed as 
\begin{equation}
\deltavec=\left({\sum A_i {\bf z}_i \over A}-\zetavec\right)\thetae,
\label{eq5}
\end{equation}
where $\zetavec$ and  ${\bf z}_i$ are the vector notations of the 
source and image positions, respectively.

Because of the very small mass ratio, planetary lensing behavior 
is well described by that of a single lens of the primary star for 
most of the event duration.  The single-lens magnification and 
centroid shift vector of the source star image are expressed 
respectively as
\begin{equation}
A = {u^2+2\over u(u^2+4)^{1/2}},
\label{eq6}
\end{equation}
\begin{equation}
\deltavec = {{\bf u} \over u^2+2}\thetae,
\label{eq7}
\end{equation}
where ${\bf u}$ is the dimensionless lens-source separation vector 
normalized by $\thetae$.  The lightcurve of a single-lens event is 
characterized by a smooth and symmetric shape.  The trajectory of 
the centroid shift traces an ellipse with a minor/major axis ratio 
of $b/a=u_0/(u_0^2+2)^{1/2}$, where $u_0$ is the impact parameter 
of the lens-source encounter normalized by $\thetae$.  Due to the 
existence of the planet, however, a short-duration perturbation can 
occur when the source star passes the region around the caustics, 
which represent the set of source positions at which the magnification 
of a point source becomes infinite.  For a planetary case, there exist 
two sets of disconnected caustics.  One small central caustic is 
located close to the host star and the other bigger planetary caustic 
is located away from the host star.  For a wide-separation planet, 
the planetary caustic has an asterisk shape with four cusps, where 
two of them are located on the star-planet axis and the other two 
are off the axis.  The center of the planetary caustic is located at  
\begin{equation}
{\bf r}_{\rm c} = {\bf s}\left( 1-{1\over s^2}\right),
\label{eq8}
\end{equation}
where ${\bf s}$ is the position vector of the planet from the star 
normalized by the Einstein radius.  The caustic size, which is 
directly proportional to the  cross-section of the planetary 
perturbation, as measured by the full width along the star-planet 
axis and a height normal to the star-planet axis are 
\begin{equation}
\Delta\xi_{\rm c} \simeq {4q^{1/2}\over s\sqrt{s^2-1}},\qquad
\Delta\eta_{\rm c} \simeq {4q^{1/2}\over s\sqrt{s^2+1}},
\label{eq9}
\end{equation}
where $q$ is the planet/star mass ratio.  For details about the 
location, shape, and size of the caustic, see \citet{han06}.  Then, 
as the separation between the star and planet increases, the position 
of the caustic moves toward the position of the planet itself 
(${\bf r}_{\rm c}\rightarrow {\bf s}$) and the size of the caustic 
rapidly decreases as $\Delta\xi\ (\Delta\eta)\propto s^{-2}$.
As a result, a wide-separation planet ($s\gg 1$) behaves as if it
is an independent lens.  Therefore, the lightcurve of an event 
with source trajectory passing through the effective region of the 
wide-separation planet appears to be that of a single-lens event
produced by a free-floating planet.

\section{Previous Methods}

There are several methods that were proposed to identify the 
existence of the primary star in isolated events caused by 
wide-separation planets, and thus enabling discrimination between 
bound and free-floating planets.  The first method is detecting 
the signature of the primary stars near the peak of lightcurves.  
The signature is produced by the planetary caustic, which is located 
at the center of the effective lensing position of the planet.  
With this method, it was estimated that primary signatures with 
$\gtrsim 5\%$ photometric deviation can be detected for $\gtrsim 
80\%$ of isolated events (with source trajectories approaching 
within the Einstein ring of the planet, $\theta_{\rm E,p}=q^{1/2}
\thetae$) caused by Jupiter-mass planets with projected separations 
$r_\perp\lesssim 10$ AU \citep{han03}.  However, the caustic shrinks 
rapidly with the increase of the planet separation as $\propto s^{-2}$, 
and thus the chance to detect the primary signature also decreases 
accordingly.  Therefore, considering that nearly one third of planets 
in our solar system have separations of $\gtrsim 20$ AU, this method 
is significantly incomplete in distinguishing between bound and 
free-floating planets.

Second, wide-separation planetary events can also be distinguished 
by the additional long-term bumps in the lightcurve caused by the 
primary star. Compared to the planetary Einstein ring, the Einstein 
ring of the primary star is much larger, and thus the source 
trajectory passing through the region around the planet has a good 
chance to approach the effective lensing region of the primary.  
\citet{han05} estimated that with this method the existence of the 
primary can be identified for $\gtrsim 50\%$ of events with 
$r_\perp \lesssim 20~{\rm AU}$.  However, the chance to detect the 
bump decreases linearly with the increase of the planetary separation, 
and thus this method is also incomplete.

The third method of identifying a wide-separation planet is detecting 
blended light from the host star.  This method is possible because 
photometry of the proposed space-based microlensing mission will not  
be affected by the blended flux from stars located close to the lensed 
source star thanks to high resolution from space observation combined 
with the fact that the main target source stars of the mission are 
low-luminosity main-sequence stars for which the lens/source flux 
ratio is relatively high.  According to \citet{bennett02}, for 
$\sim 1/3$ of events with detected planets from a space lensing 
mission, the planetary host star is either brighter than or within 
$\sim 2$ magnitudes of the source star's brightness.  This method has 
an advantage that primary detection is possible regardless of the 
planetary separation, but it has a disadvantage that it cannot be 
applicable to events associated with faint or dark primaries.

\section{New Method: Astrometric Follow-up}

In this section, we demonstrate that astrometric follow-up observations 
of isolated planetary lensing events by using high-precision 
interferometers can firmly break the degeneracy in the events produced 
by wide-separation and free-floating planets.  The method is based on 
the fact that astrometric lensing effect covers much longer range of 
lens-source separation than photometric effect \citep{miraldaescude96}.  
In the limiting case of $u\gg 1$, the astrometric and photometric effects 
are approximated respectively as
\begin{equation}
A \rightarrow 1+{2\over u^4},
\label{eq10}
\end{equation}
\begin{equation}
\delta \rightarrow {\thetae\over u}.
\label{eq11}
\end{equation}
Therefore, the photometric effect vanishes rapidly as $\propto u^{-4}$, 
while the astrometric effect decays much more slowly as $\propto u^{-1}$.   
Then, for an isolated event produced by a wide-separation planet, the 
astrometric effect of the primary will be considerable although its
photometric effect is negligible.

\begin{figure}[t]
\epsscale{1.15}
\plotone{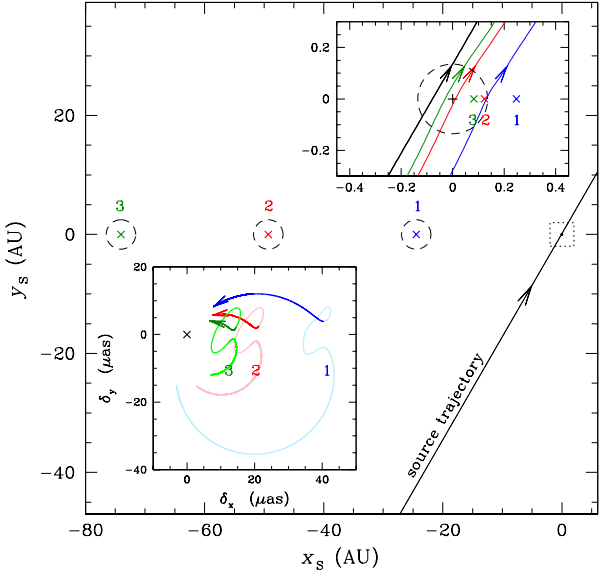}
\caption{\label{fig:one}
The geometry of isolated 
planetary events produced by wide-separation planets with various 
distances from the primary (main panel) and the resulting trajectories 
of the source star image centroid shifts (lower left inset).  The 
inset on the upper right side shows the 
enlarged view of the region around the planet [corresponding to the 
region inside the dotted square centered on (0,0) in the main panel]. 
See detailed explanation in text.
}\end{figure}

In Figure~\ref{fig:one}, we illustrate the basic scheme of the method.
In the main panel of the figure, we present the geometry of isolated 
planetary events produced by wide-separation planets with various 
distances from the primary.  The coordinates are centered at the 
effective position of the planet.  The effective planet position is 
located at a position with a separation $1/s$ from the planet toward 
the primary and it corresponds to the center of the planetary caustic.  
For the physical parameters of the lens system, we assume that the mass 
of the lens is $m=0.5\ M_\odot$ and the distances to the lens and source 
stars are $D_{\rm L}=6$ kpc and $D_{\rm S}=8$ kpc, respectively, by 
adopting the values of a typical Galactic bulge event.  We assume 
that the planet/primary mass ratio is $q=3\times 10^{-3}$.  The 
primaries are located on the left side of the planet with normalized 
separations of $s=10$, 20, and 30.  In physical units, these separations 
correspond to the projected separations of $r_\perp = 24.7$ AU, 49.4 AU, 
and 74.1 AU, respectively.  The positions of the individual primary 
stars are marked by crosses with blue (also marked by number `1'), 
red (`2'), and green (`3') colors and the dashed circles around the 
individual primaries represent the Einstein rings of the primaries.  
The straight line with an arrow represents the source trajectory.  
The inset on the upper right side shows the enlarged view of the 
region around the planet [corresponding to the region inside the 
dotted square centered on (0,0) in the main panel], where the dashed 
circle is the Einstein ring of the planet, the black straight line is 
the source trajectory, and the crosses and curves drawn in colors  
represent the true positions of the planets and the trajectories of 
the source star image centroid for the three cases of lens system 
geometry with primaries located at the positions marked by the 
corresponding colors (and numbers).  The inset on the lower left 
side shows the trajectories of the centroid shifts of the events 
produced by the individual primary-planet pairs.  Each trajectory 
is drawn in two tones of color, where the part drawn in the dark-tone 
color indicates the trajectory after the planetary lensing event, 
while the light-tone part indicates the trajectory before and during 
the event.  Since astrometric follow-up observations will be carried 
out after the event, the dark-tone part of the trajectory is what will 
be observed.

From the figure, we find the following trends of centroid shifts.
\begin{enumerate}
\item
Due to the much larger size of the Einstein ring of the primary, 
the centroid motion is dominated by the astrometric effect of the 
primary in most of the time of the centroid motion.  Since the 
source trajectory has a large impact parameter with respect to 
the primary, i.e.\ $u_0\gg 1$, the minor/major axis ratio of the 
primary-induced centroid shift trajectory is 
$b/a=u_0/(u_0^2+2)^{1/2}\rightarrow 1$, and thus the trajectory has 
a circular shape.
\item
The astrometric effect of the planet becomes equivalent to or larger 
than that of the primary in the region where $d_{\perp,{\rm p}}
\lesssim q d_{\perp,\star}$.  Here $d_{\perp,{\rm p}}$ and 
$d_{\perp,\star}$ represent the projected distances from the source 
to the planet and primary star, respectively.  As a result, during 
the time when the source is located within this region, the centroid 
shift trajectory deviates from the circular one induced by the primary.
However, due to the small mass ratio of the planet, the deviation 
lasts only a short period of time.  The planet-induced perturbation 
traces an elliptical trajectory with an axis ratio of $u_{0,{\rm p}}
/(u_{0,{\rm p}}^2+2)^{1/2}$, where  $u_{0,{\rm p}}$ is the planet-source 
impact parameter normalized by the Einstein radius of the planet.
Therefore, if the source passes closer to (further from) the effective
planet position than the case shown in Figure~\ref{fig:one}, the part
of the elliptical trajectory induced by the planet would be more 
elongated (circular).  However, this shift of the source trajectory 
has little effect on the source-primary separation, and thus the 
part of the centroid shift trajectory induced by the primary remains 
nearly the same.
\item
Despite the large separation between the primary and the source near 
the planet, the centroid shift induced by the primary is substantial.
The amount of the primary-induced centroid shift is 
\begin{equation}
\delta \simeq 40\ \mu{\rm as}\ \left( {\thetae\over 410\ \mu{\rm as}}
\right) \left( {r_\perp\over 10\ {\rm AU}} \right)^{-1}. 
\label{eq12}
\end{equation}
\end{enumerate}

\begin{figure}[t]
\epsscale{1.15}
\plotone{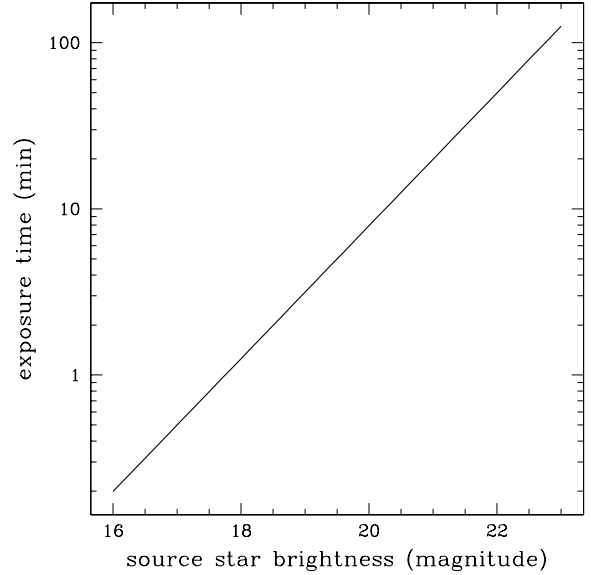}
\caption{\label{fig:two}
The {\it SIM} exposure time required to achieve positional accuracy 
better than a threshold value $\delta_{\rm th}=4~\mu{\rm as}$ as a 
function of source brightness.  The threshold value corresponds to 
the amount of the centroid shift caused by a primary star of a 
wide-separation planetary system with a projected separation of 
$r_\perp=100~{\rm AU}$ from the planet.
}\end{figure}

Then, a question is how well the astrometric signature of primaries 
can be measured by the {\it SIM} observations.  According to the 
specification of the {\it SIM}, it will have positional accuracy of 
$\sigma\sim 0.4~\mu{\rm as}$ for a star of $V=12$ with an exposure 
time of $t_{\rm exp}=0.5~{\rm min}$ and the uncertainty increases with 
the decrease of the photon count.  According to equation~(\ref{eq12}), 
the threshold centroid shift to identify the existence of primaries 
with separations up to $r_\perp\sim 100~{\rm AU}$ is $\delta_{\rm th}
\sim 4~\mu{\rm as}$.  Then, the exposure time required to achieve 
astrometric accuracy to measure the threshold centroid shift is 
\begin{equation}
t_{\rm exp} \simeq
0.08~{\rm min} \left( {\delta_{\rm th}\over 4~\mu{\rm as}}
\right)^{-2} 10^{0.4(V-12)},
\label{eq13}
\end{equation}
where $V$ is the source star magnitude.  In Figure~\ref{fig:two}, 
we present the required exposure time as a function of source star 
brightness.  The exposure time in equation~(\ref{eq13}) is based 
on photon noise.  However, even considering other sources of 
noise, it will be possible to measure centroid shift of $\delta > 
\delta_{\rm th}$ for events involved with source stars brighter than 
$V\sim 19$ with exposure times of $t_{\rm exp}\lesssim 10~{\rm min}$.  
Therefore, by focusing on isolated events associated with bright 
source stars, astrometric follow-up observations can be carried 
out not seriously affecting the {\it SIM} lensing observations for 
its original goal.  In addition, the centroid motion is slow when 
the source is away from the primary, and thus follow-up does not 
require prompt response just after the planetary lensing.

Compared to the previous astrometric methods, the proposed method 
has the following important advantages.  First, the proposed method 
is complete in the sense that  it can identify the bound nature of 
nearly all planets within the possible range of distance ($\lesssim 
100~{\rm AU}$) from the primary.  Second, the astrometric effect of 
the primary does not depend on brightness of the primary, and thus 
the proposed method can be applied to planets accompanied with faint 
or even dark primaries.  
%
%
Third, astrometric observation of the primary-induced 
centroid shift enables one to measure the angular Einstein 
ring radius.  While the Einstein timescale, which is 
determined from photometric observation, results from 
the combination of the three physical parameters of the lens
mass, distance to the lens, and lens-source 
transverse speed, $\theta_{\rm E}$ does not depend on the 
transverse speed.
Therefore, one can better constrain the mass of the primary,
and thus that of the planet.

\section{Conclusion}

We showed that free-floating 
planets can be securely identified by conducting astrometric follow-up 
observations of isolated events to be detected in future photometric 
lensing surveys by using high-precision interferometers to be operated 
contemporarily with the photometric surveys.  The method is based on 
the fact that astrometric lensing effect covers much longer range of 
the lens-source separation than the photometric effect.  We demonstrated 
that several astrometric follow-up observations of isolated planetary 
events associated with source stars brighter than $V\sim 19$  by using 
the {\it Space Interferometry Mission} with an exposure time of 
$\lesssim 10~{\rm min}$ for each observation will make it possible 
to measure the centroid shift induced by primaries with projected 
separations up to $\sim 100~{\rm AU}$.  Therefore, the proposed method 
is far more complete than previously proposed methods that are flawed 
by the limited applicability only to planets with projected separations 
$\lesssim 20~{\rm AU}$ or planets accompanied by bright primaries.

\acknowledgments
This work was supported by the Astrophysical Research Center for 
the Structure and Evolution of the Cosmos (ARCSEC) of the Korea 
Science \& Engineering Foundation (KOSEF) through the Science 
Research Program (SRC) program.

\end{document}